\documentclass[11pt]{article}
\usepackage{moriond,epsfig}

\def\be{\begin{equation}}
\def\ee{\end{equation}}
\def\bea{\begin{eqnarray}}
\def\eea{\end{eqnarray}}

\def\de   {\Delta E}
\def\mes  {m_{\mbox{\scriptsize ES}} }

\usepackage{relsize}
\def\babar{\mbox{\slshape B\kern-0.1em{\smaller A}\kern-0.1em
    B\kern-0.1em{\smaller A\kern-0.2em R}}}

\begin{document}
\vspace*{4cm}
\title{Studies of Rare Hadronic B Decays with \babar\ }

\author{ J. KROSEBERG }

\address{Santa Cruz Institute for Particle Physics, University of California Santa Cruz,\\
1156 High Street, Santa Cruz, CA 95062, USA}

\maketitle\abstracts{
We present a selection of recent results from studies of rare hadronic $B$ decays based on 
a sample of 232 million $B\overline{B}$ pairs, corresponding to an integrated 
luminosity of 211 fb$^{-1}$, which were recorded with the \babar\ detector at the 
PEP-II asymmetric-energy $e^+e^-$  storage ring. 
}

\section{Introduction} 

\vspace*{-1.5ex}
The study of rare hadronic $B$ decays provides valuable insight into various aspects 
of particle interactions. The large \babar\ data set makes it possible to probe branching 
fractions (BF) down to $\mathcal{O}(10^{-6})$. Within the standard model of particle physics 
(SM), these measurements test descriptions of specific decay mechanisms as well as more 
fundamental concepts such as QCD factorization. Meaningful constraints can be obtained for 
theories beyond the SM using processes where the SM predictions are significantly lower
than current experimental sensitivity. Another important aspect is to probe our understanding 
of the CKM description of quark mixing and CP violation; this, however, is covered in a 
different contribution to this conference.~\cite{laumoriond}

\section{Experimental strategy}

\vspace*{-1.5ex}
The \babar\ detector is described elsewhere.\cite{detector} Key selection quantities are the 
missing energy $\de \equiv E^*_{B}-E_{\rm beam}^*$ and the beam-energy-substituted mass 
$\mes \equiv \sqrt{ E^{*2}_{\rm beam}-{\vec{p}}_{B}^{\;*2}}$, where $E_{\rm beam}^*$    
denotes the center-of-mass (CM) beam energy; $E^*_B$ and ${\vec{p}_{B}^{\;*}}$ 
are the reconstructed $B$ mass and momentum in the CM frame. After reducing the large backgrounds, 
for example by applying event shape requirements to discriminate  between the jet-like continuum 
background (\mbox{$e^+e^-\to q\bar q$}, where $q=u,d,s,c$) and the more spherically-symmetric 
signal events, the signal is extracted using extended maximum-likelihood fits to the data.

\vfill
  
\newpage

\section{Decays to open-charm mesons}

\vspace*{-1.5ex}
$B$-meson decays to double-charm final states $D^{(*)}\overline{D}^{(*)}$ are a 
good example for the wide range of questions which can be addressed through the 
experimental study of rare hadronic $B$ decays. In addition to constraining the 
SM CKM triangle, measurements of these processes provide a test of QCD heavy-heavy 
factorization as well as information on the decay mechanism and final-state interactions. 
The measured BF from a recent analysis~\cite{DD} are summarized in Table~\ref{tab:BDD}. 
These include BF limits from a first search for the color-suppressed $B^0\to D^0\overline{D}^0$, 
which -- if observed -- would provide evidence of $W$-exchange or annihilation contributions.
 \begin{table}[t]
 \caption{Measured $B\to D^{(*)}\overline{D}^{(*)}$ branching fractions. \label{tab:BDD}}
 \begin{center}
 \begin{tabular}{|l|c|c|c|c|c|}
 \hline
 {\small Mode} &{\small $B^0\to D^0\overline{D}^0$}  &{\small $B^0\to D^{*0}\overline{D}^{0}$} 
 &{\small $B^0\to D^{*0}\overline{D}^{*0}$} &{\small $B^0\to D^+D^-$} 
 &{\small $B^0\to D^{*+}D^{-}$} \\
 {\small BF}/{\tiny$10^{-4}$} & {\footnotesize$<0.59$}  &{\footnotesize$<2.92$}  & {\footnotesize$<0.92$} 
 &{\footnotesize $2.81\pm0.43\pm0.45$} & {\footnotesize $5.72\pm0.64\pm0.71$}\\
\hline
 {\small Mode}&{\small $B^0\to D^{*+}D^{*-}$} &{\small $B^-\to D^{-}\overline{D}^{0}$} 
 &{\small $B^-\to D^{-}\overline{D}^{*0}$} &{\small  $B^-\to D^{*-}\overline{D}^{*0}$} 
 &{\small $B^-\to D^{*-}\overline{D}^{*0}$}  \\
 {\small BF}/{\tiny $10^{-4}$}           & {\footnotesize $8.11\pm0.57\pm0.99$} 
 &{\footnotesize $3.76\pm0.57\pm0.45$}  & {\footnotesize $3.56\pm0.52\pm0.40$} 
 &{\footnotesize $6.30\pm1.32\pm0.95$}  & {\footnotesize $8.14\pm1.17\pm1.21$} \\
\hline
 \end{tabular}
 \end{center}
 \end{table}

The $W$-exchange contribution to $B^0\to D^{*+}D^{*-}$ decays could be estimated from a 
measurement of  the relative rate with respect to the decays $B\to D^{(*)-}_s\overline{D}^{(*)+}_s$, 
which in the SM are dominated by $W$-exchange. These have been searched for~\cite{DsDs} 
for the first time, without observing a significant signal. The resulting 90\% C.L. 
upper BF limits, $1.0\times10^{-4}$ ($D_s\overline{D}_s$),  $1.3\times10^{-4}$ 
($D^{*-}_s\overline{D}_s$), and $2.4\times10^{-4}$  ($D^{*}_s\overline{D}^{*}_s$), 
are close to the expectation from perturbative QCD calculations and are already excluding 
certain alternative models.

\begin{figure}[h]
\vspace*{-0mm}
\psfig{figure=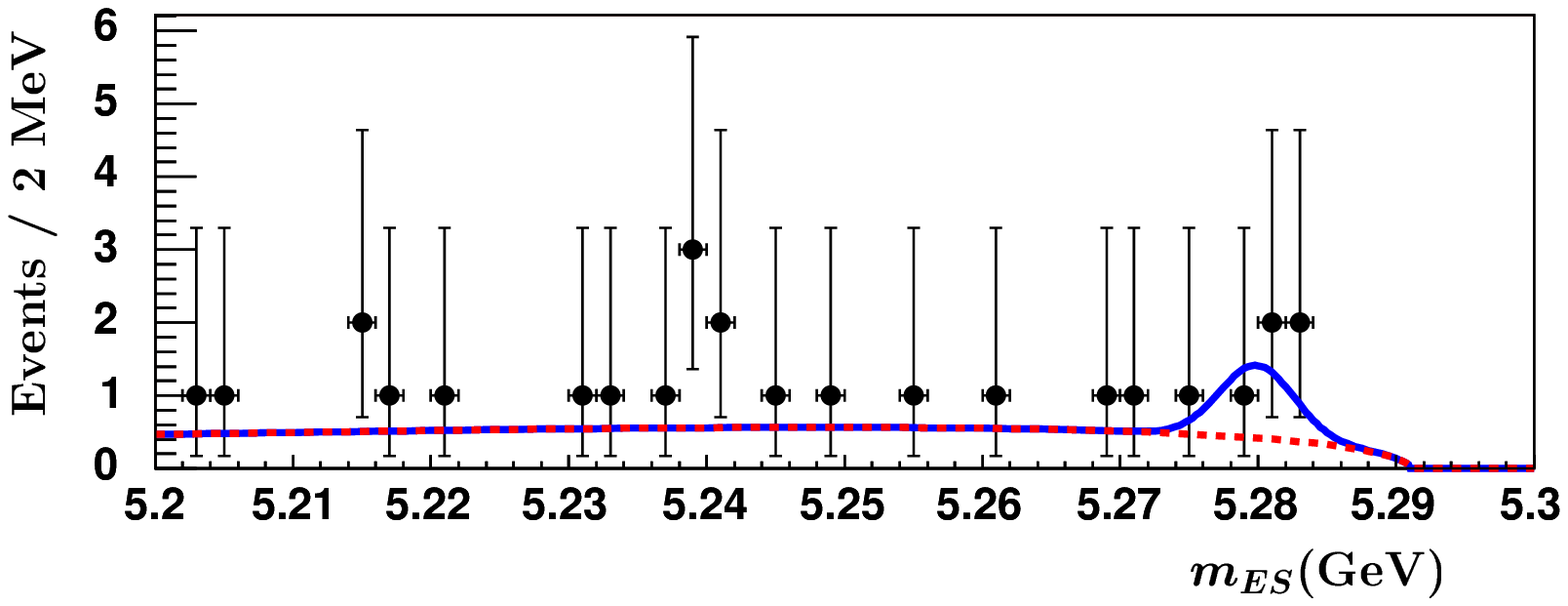,height=1.43in}
\psfig{figure=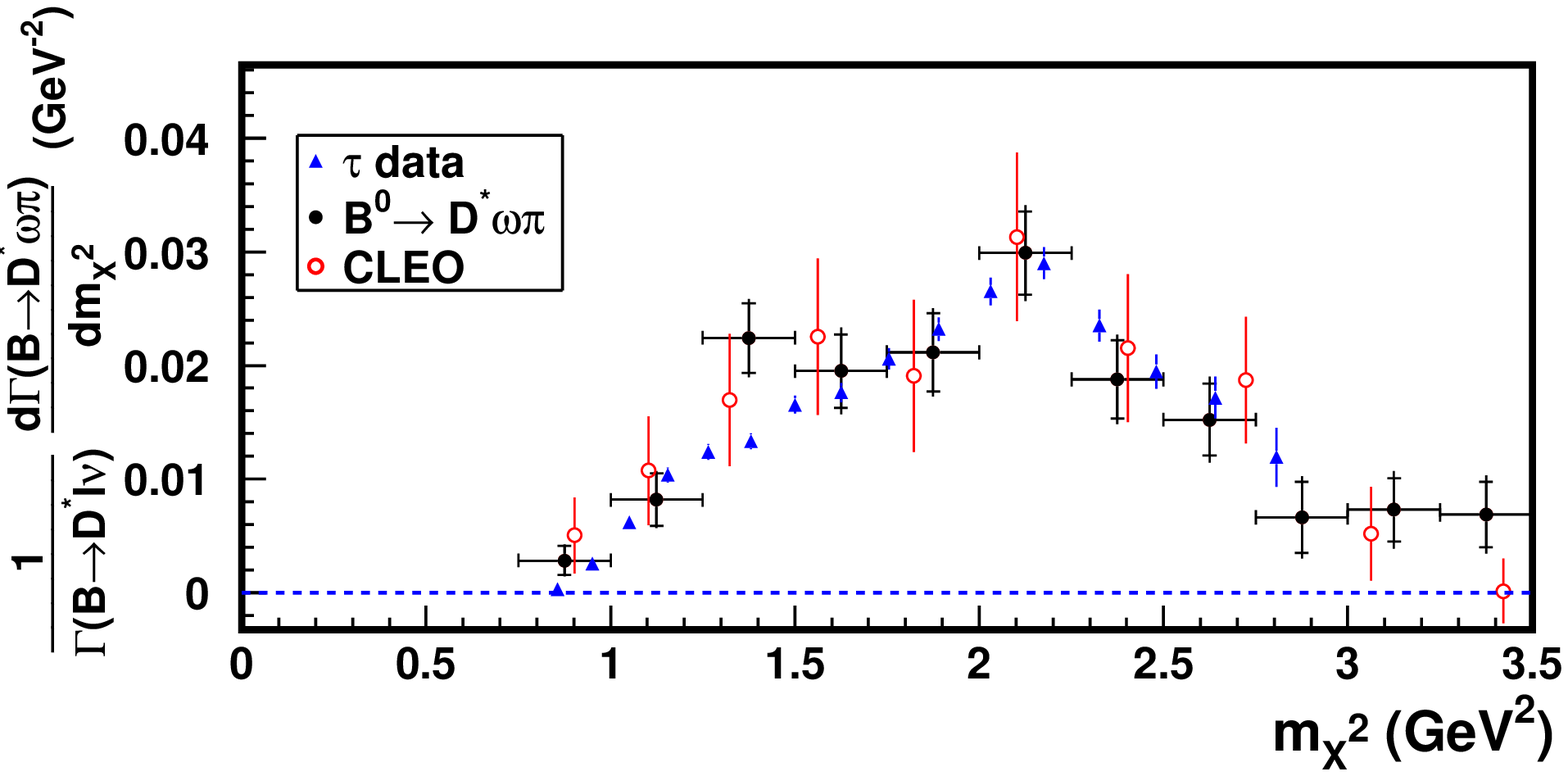,height=1.45in}

\setlength{\unitlength}{1cm}    
\begin{picture}(0.1,0.1)
\put(0.4,4.0){(a)}
\put(9.5,4.0){(b)}
\end{picture}
\vspace*{-7mm}
\caption{(a) $\mes$ distribution for $B^-\to D^{*-}_s\phi$ candidates in the $\de$ signal region; (b) $m^2_{X}$ distribution 
for $B \to D^{*}\omega\pi$ candidates, normalized to the semi-leptonic width $\Gamma(\overline{B}^0\to D^{*+}l^-\bar{\nu})$, 
compared to other measurements.
\label{fig:dsphi}}
\end{figure}

\babar\ also performed a search~\cite{DsPhi} for the annihilation processes 
$B^-\to D^{(*)-}_s\phi$, see Figure~\ref{fig:dsphi}(a), yielding upper BF limits 
of $1.9\times10^{-6}$ and $1.2\times10^{-5}$ for $B^-\to D^{-}_s\phi$ and 
$B^-\to D^{*-}_s\phi$ respectively. Although still about a factor of ten larger 
the SM expectation, this constitutes an improvement of previous limits by two 
orders of magnitude and makes it possible to set competitive limits on parameters 
in the type II Two-Higgs-Doublet Model and the MSSM with $R$-parity violation.

An elegant test of factorization is provided in the context of 
a study of $B \to D^{*}\omega\pi$ decays~\cite{DOmPi}, where the invariant mass 
spectrum of the $\omega \pi$ system is found to be in agreement with 
theoretical expectations based on $\tau$ decay data, cf. Figure~\ref{fig:dsphi}(b).
In addition, the longitudinal polarization fraction is measured from a $D^*$ 
angular analysis and found to be consistent with heavy-quark effective theory 
predictions assuming factorization and using results from semi-leptonic $B$ decays.

\section{Inclusive decays to open charm}

\vspace*{-1.5ex}
An inclusive picture of $B$ decays to open charm is provided by a preliminary measurement~\cite{recoilC} 
of ${D^0 X}$, ${D^+ X}$, ${D^+_s X}$, ${\Lambda^+_c X}$ final states and their charge conjugates, which 
is performed separately for flavor-correlated and anti-correlated $b$ to $c$ transitions. Events are 
selected by completely reconstructing one $B$ and searching for a reconstructed charm particle in the 
rest of the event. 

The measured branching fractions include BF$(B^-\to D^+_s X)=1.1 ^{+0.4}_{-0.3} \pm0.1^{+0.2}_{-0.1}$, 
which is the first evidence of $D^+_s$ production in $B^-$ decays. The sum of all correlated charm 
branching fractions is found to be compatible with 1. The numbers of charm particles per $B^-$ decay  
($n_c^- = 1.202 \pm 0.023 \pm 0.040 ^{+0.035}_{-0.029} $) and per $\bar{B}$ decay ($n_c^0 = 1.193\pm 
0.030\pm 0.034^{+0.044}_{-0.035}$) are consistent with previous measurements and theoretical expectations.
Charm momentum distributions in the $B$ rest frame are measured as well; two examples are shown in 
Figs.~\ref{fig:charmcount}(a) and (b).

\section{Baryonic decays}

\begin{figure}[hb]
\vspace*{2mm}
\psfig{figure=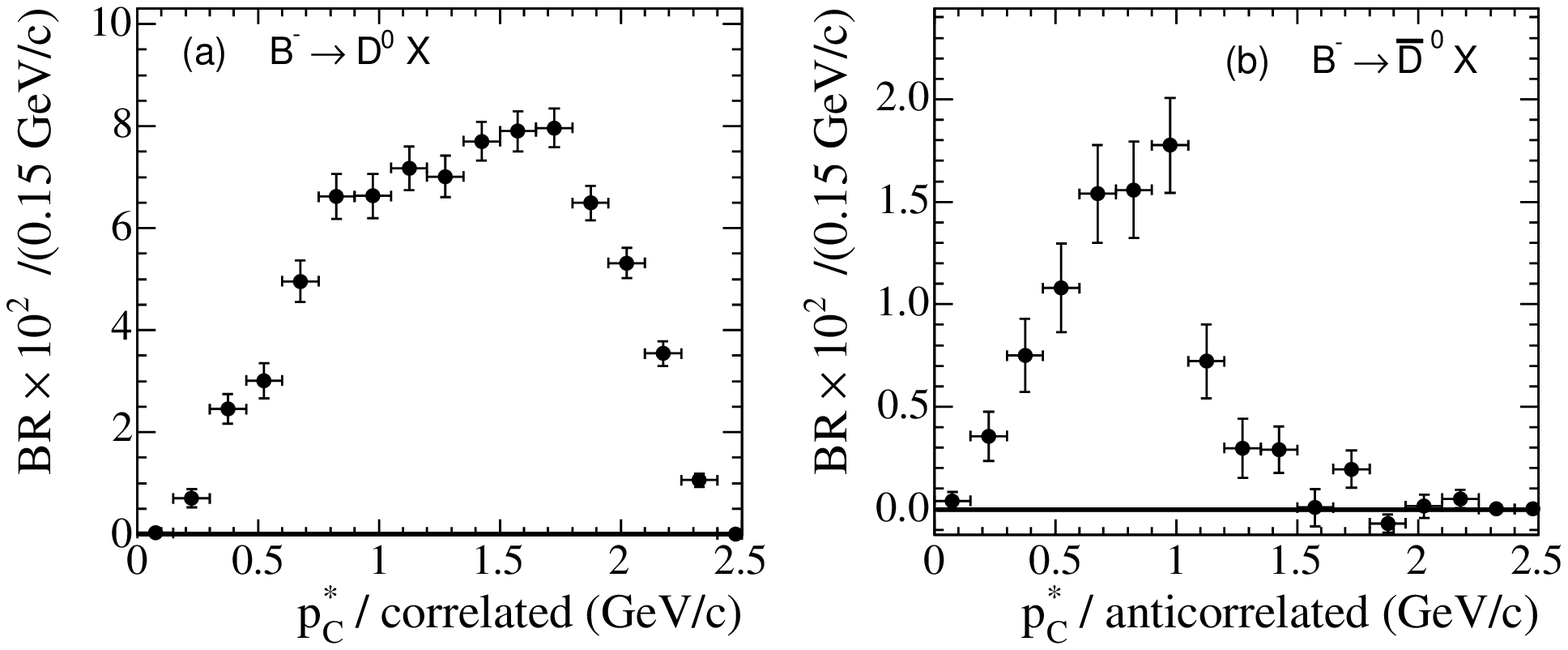,height=1.2in}
\hspace*{-2mm}\psfig{figure=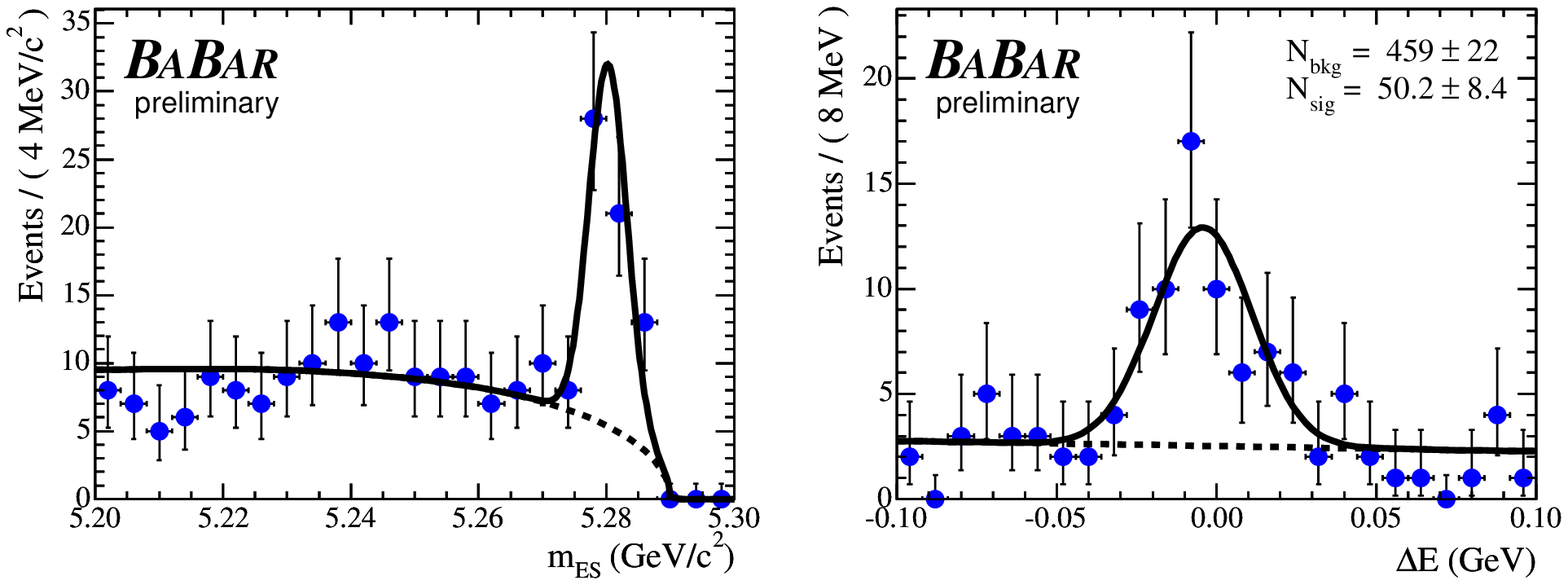,height=1.23in,clip=true}
\psfig{figure=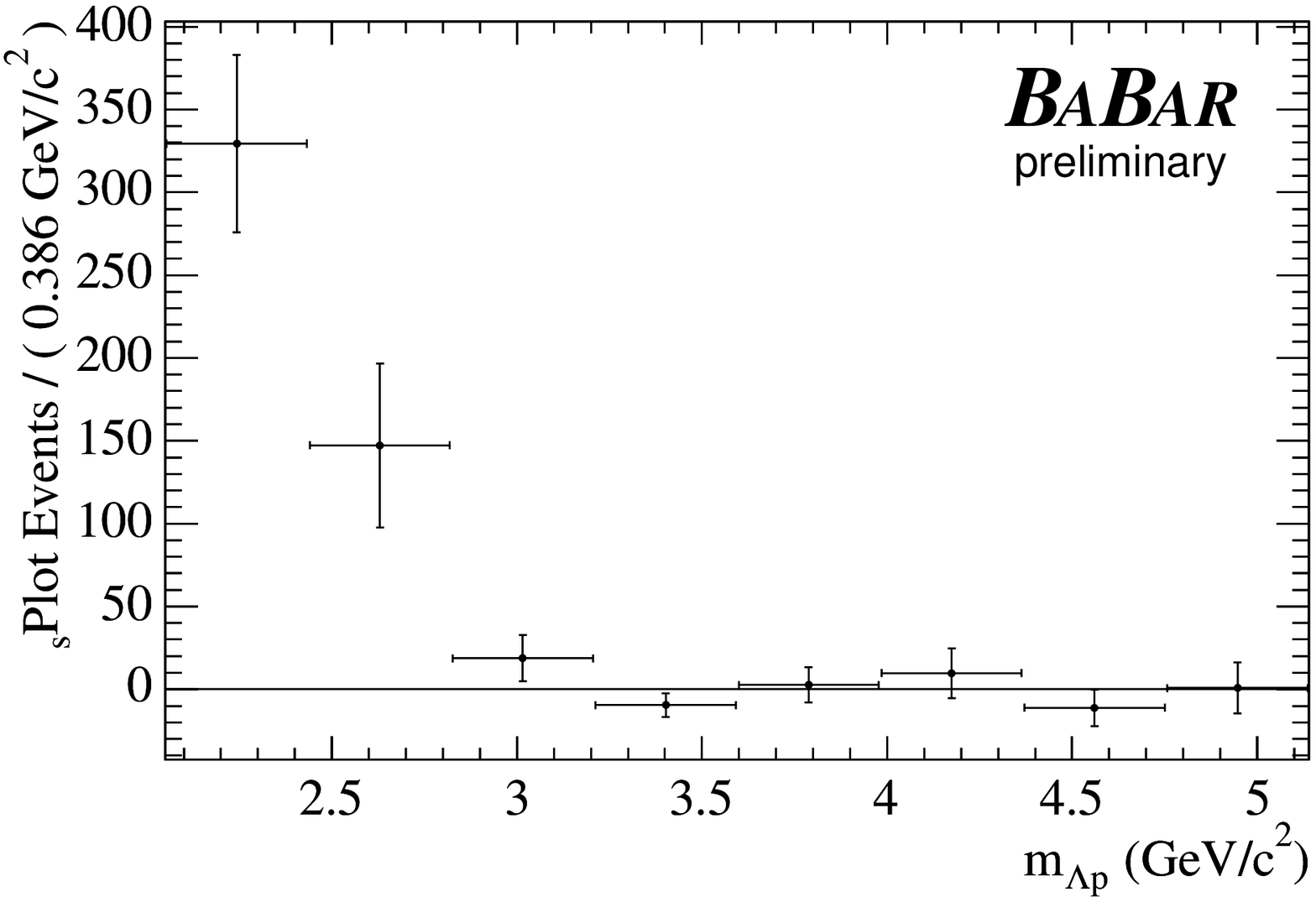,height=1.23in}

\setlength{\unitlength}{1cm}    
\begin{picture}(0.1,0.1)
\put(0.4,3.7){(a)}
\put(4.1,3.7){(b)}
\put(7.7,3.7){(c)}
\put(11.7,3.7){(d)}
\put(1.5,1.3){\psfig{figure=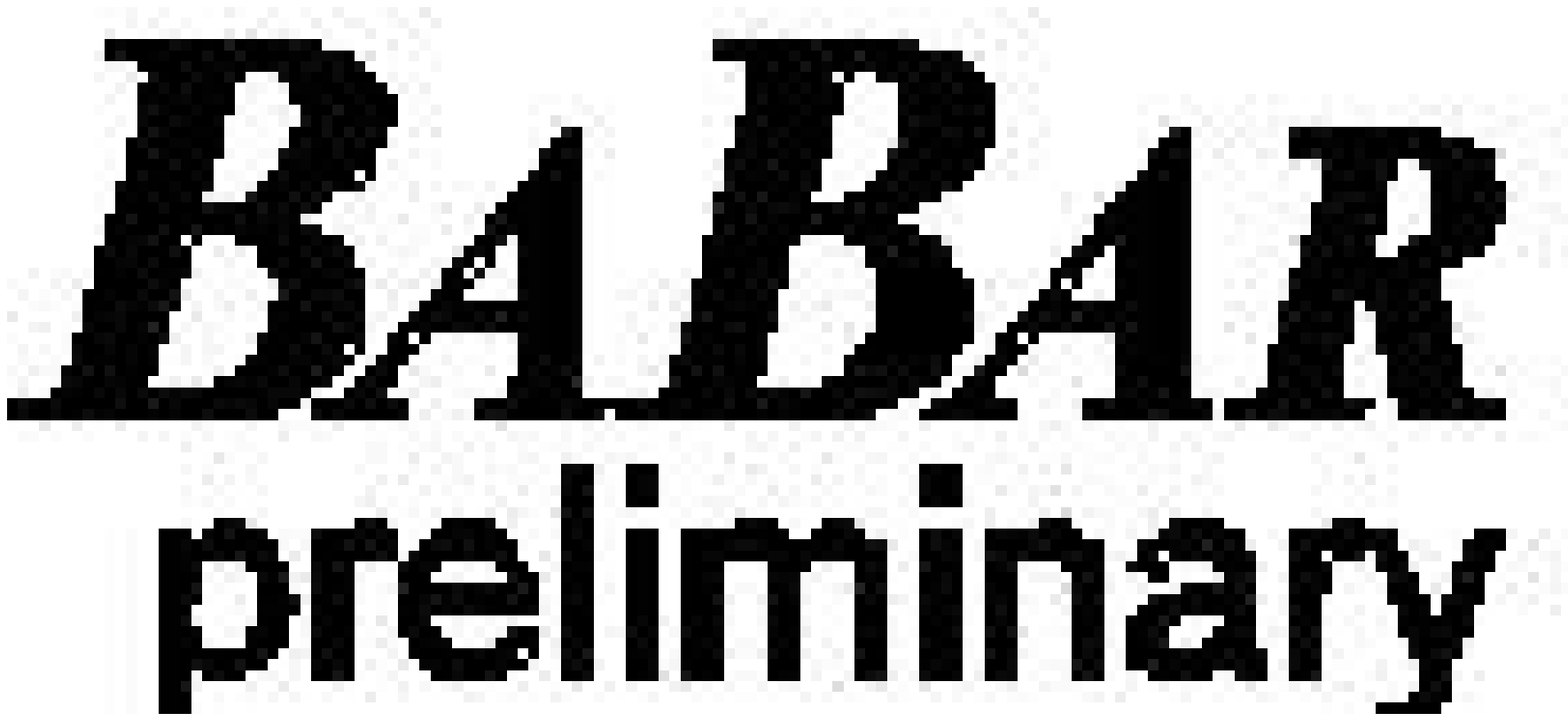,height=0.16in}}
\put(5.8,2.3){\psfig{figure=bbprelim.eps,height=0.16in}}
\end{picture}
\vspace*{-7mm}
\caption{Charm momentum distributions in the $B$ rest frame for (a) $B^-\to D^0X$ and (b) $B^-\to \overline{D}^0X$ decays;    
(c) $\mes$ distribution for selected $\overline{B}^0 \to \Lambda_c \bar{p}$ candidates; 
(d) $m(\Lambda p)$ distribution for $B^0\rightarrow \bar{\Lambda} p \pi^-$ candidates.\hfill \
\label{fig:charmcount}}
\end{figure}

\vspace*{-1.5ex}
Two-body decays to charm baryons provide a way to test various theoretical models 
for exclusive baryonic $B$ decays. In particular, there is theoretical interest in 
the suppression of the two-body baryonic decay rates compared to three-body rates 
and the possible connection to the mechanisms for baryon production in $B$ decays. 
Figure~\ref{fig:charmcount}(c) shows the $\mes$ projection of of selected   
$\overline{B}^0 \to \Lambda_c \bar{p}$ signal events. The corresponding BF is measured 
to be $(2.15 \pm 0.36 \pm 0.13 \pm 0.56) \times 10^{-5}$, where the third uncertainty 
is associated with the $\Lambda_c\to pK^-\pi^+$ BF.

Charmless two-body baryonic $B$ decays have not yet been seen but several observation 
of three-body decays have been reported. A recent example is the \babar\ measurement 
of the $B^0\rightarrow \bar{\Lambda} p \pi^-$ BF yielding  
$\left({3.34}\pm{0.54}\pm{0.31}\right)\times 10^{-6}$. The efficiency-corrected distribution 
of the di-baryon invariant mass $m(\Lambda p)$, see Figure~\ref{fig:charmcount}(d), 
shows a near-threshold enhancement already seen in several other baryonic B decays. 

All these results, which are preliminary, are consistent with previous measurements 
by the Belle collaboration.

\section{Decays to charmless mesons}

\vspace*{-1.5ex}
Tables~\ref{tab:cless}(a) and (b) summarize recent results of searches for $B$ decays 
to charmless mesons. The study of such processes is a rich field and a 
comprehensive review is far beyond the scope of this note. One of various 
motivations arises from  potential differences $\Delta S$ between the measurements of 
the CKM parameter $\sin2\beta$ from tree and penguin-dominated processes 
respectively. ~\footnote{
Although below most of the results will be discussed  
in the context of SM constraints on  $\Delta  S$, it should be noted that 
these measurements are very interesting in their own rights; see the references for details.}
Leading order SM calculations give $\Delta S=0$ and significant 
non-zero values could be a signal for physics beyond the SM. There are, however, 
also contributions from sub-leading SM amplitudes, often referred to as SM 
pollution, which involve the CKM matrix element $V_{ub}$ and are currently not well known.

Using SU(3) flavor symmetry, a SM constraint on $\Delta S_{\phi K^0}$, i.e.\ the SM pollution 
for the decay $B^0\to\phi K^0$, can be obtained  from the BF of eleven different $B^0$ 
decays.\cite{SU3} Experimental limits for nine of these BF were previously available; a 
recent \babar\ search~\cite{KKs} for $B^0\to K^{*0}K_S$ completes the set. The resulting 
limit for the sum of branching fractions BF($B^0\to \overline{K}^{*0}K^0$)+
\mbox{BF($B^0\to {K}^{*0}\overline{K}^0$)}, see Table~\ref{tab:cless}(a), makes it possible 
to set a 90\% C.L. upper bound on $\Delta S_{\phi K^0}$ of 0.43. Recently improved BF 
limits~\cite{PhiPi} for the decay $B^0\to\phi\pi^0$, which is also relevant for constraining  
$\Delta S_{\phi K^0}$, and its charged counterpart  $B^+\to\phi\pi^+$ are included in 
Table~\ref{tab:cless}(a) as well.

The contribution from SM pollution to the process $B^0\to\eta' K_S$ can be constrained in a similar 
way\cite{SU3}. New BF limits~\cite{EtaPi} for three of the relevant decays ($B^0\to\eta'\eta$, 
$B^0\to\eta\pi^0$, and $B^0\to\eta'\pi^0$), as listed in Table~\ref{tab:cless}(a), represent two 
to three-fold improvements over previous measurements. The corresponding reduction in 
the theoretical uncertainty on $\Delta S_{\eta' K_S}$ is estimated to be $20\%$.

$B^0$ decays to the $CP$ eigenstate $K^0_S K^0_S K^0_L$ are pure $b\to s \bar s s$ penguin 
transitions. $b\to u$ decay amplitudes enter only through rescattering, which reduces the 
corresponding SM $CP$ asymmetry uncertainty. In a first search~\cite{KsKsKl} for these decays, 
\babar\ finds no significant signal and obtains the results listed in Table~\ref{tab:cless}(b). 
Here, the central BF value is determined assuming a uniform 3-body phase space and excluding 
the $\phi$ resonance; the upper limit is single-sided Bayesian and based on a uniform prior 
probability for physical values.

In addition, \babar\ studied the decays of charged and neutral $B$ mesons to $\eta'K^*$ and $\eta'\rho$, 
resulting in the observation of 
$B^0\to\eta'K^{*0}$, evidence for $B^+\to\eta'K^{*+}$, and significantly improved limits for the 
other modes, see Table~\ref{tab:cless}(b).

\renewcommand{\arraystretch}{1.1}
 \begin{table}[t]
 \caption{ Recent results of searches for $B$ decays to charmless mesons. Results marked ${}^\dagger$ are preliminary.\hfill\ \label{tab:cless}}

\vspace*{3.0ex}
\begin{minipage}[b]{0.5\textwidth}
 \begin{tabular}{|l|cc|}
 \hline
{\small \ \hfill Mode\hfill\ }   & {\ \hfill\small BF/$10^{-6}$\hfill\ }  & {\small \ \hfill Limit {\tiny (90\% C.L.)}\hfill\ } \\\hline\hline
{\small $B^0\to K^{*0}K_S$}$^\dagger$ & & \\
{\tiny($\overline{K}^{*0}K^0+{K}^{*0}\overline{K}^0$)} &\raisebox{1.5ex}[-1.5ex]{{\small$0.2^{+0.9}_{-0.8}{}^{+0.1}_{-0.3}$}} &
         \raisebox{1.5ex}[-1.5ex]{{\small$<1.9\times10^{-6}$}} \\\hline
{\small $B^0\to\phi\pi^0$}  &{\small$0.12\pm0.13$} & {\small$<0.28\times10^{-6}$}\\
{\small $B^+\to\phi\pi^+$}  &{\small$-0.04\pm0.17$} & {\small$<0.24\times10^{-6}$} \\\hline
{\small $B^0\to\eta'\eta$}  &{\small$0.2^{+0.7}_{-0.5}{}^{+0.4}_{-0.4}$}& {\small$<1.7\times10^{-6}$} \\
{\small $B^0\to\eta\pi^0$}  &{\small$0.6^{+0.5}_{-0.4}{}^{+0.1}_{-0.1}$}& {\small$<1.3\times10^{-6}$}\\
{\small $B^0\to\eta'\pi^0$} &{\small$0.8^{+0.5}_{-0.4}{}^{+0.1}_{-0.1}$}& {\small$<2.1\times10^{-6}$}\\
\hline
 \end{tabular}
\end{minipage}
\begin{minipage}[b]{0.5\textwidth}
 \begin{tabular}{|l|ll|}
 \hline
{\small \ \hfill Mode\hfill\ }   & {\ \hfill\small BF/$10^{-6}$\hfill\ }  & {\small \ \hfill Limit {\tiny (90\% C.L.)}\hfill\ } \\\hline\hline
{\small $B^0\to K^0_S K^0_S K^0_L$}$^\dagger$&{\small$2.4^{+2.7}_{-2.5}{}^{+0.6}_{-0.6}$}         &{\small$<6.4\times10^{-6}$} \\\hline
{\small $B^0\to \eta'K^{*0}$}\hfill\ $^\dagger$   &{\small$3.8^{+1.1}_{-1.1}{}^{+0.5}_{-0.5}$} &{\small\ \hfill -- \hfill \ } \\
{\small $B^+\to \eta'K^{*+}$}\hfill\ $^\dagger$   &{\small$4.9^{+1.9}_{-1.7}{}^{+0.8}_{-0.8}$}         &{\small$<7.9\times10^{-6}$} \\
{\small $B^0\to \eta'\rho^{0}$}\hfill\ $^\dagger$ &{\small$0.4^{+1.2}_{-0.9}{}^{+1.6}_{-0.6}$} &{\small$<3.7\times10^{-6}$} \\
{\small $B^+\to \eta'\rho^{+}$}\hfill\ $^\dagger$ &{\small$6.8^{+3.2}_{-2.9}{}^{+3.9}_{-1.3}$} &{\small$<14\times10^{-6}$} \\
{\small $B^0\to \eta'f_{0}$}\hfill\ $^\dagger$    & & \\
{\small ($f_{0}\to\pi^+\pi^-$}) &\raisebox{1.5ex}[-1.5ex]{{\small$0.1^{+0.6}_{-0.4}{}^{+0.9}_{-0.4}$}} &\raisebox{1.5ex}[-1.5ex]{$<1.5\times10^{-6}$} \\
\hline
 \end{tabular}
\end{minipage}

\setlength{\unitlength}{1cm}    
\begin{picture}(0.1,0.1)
\put(0.0,4.6){(a)}
\put(8.0,4.6){(b)}
\end{picture}
 \end{table}

\section*{Acknowledgments}

\vspace*{-1.5ex}
The author gratefully acknowledges the support provided by the US Department of Energy,  
the National Science Foundation,  
and the Marie Curie Actions program of the European Union.

\end{document}